\documentclass[%
 reprint,
superscriptaddress,longbibliography,
 amsmath,amssymb,aps, prb, floatfix, showpacs
]{revtex4-2}

\usepackage{graphicx}
\usepackage{dcolumn}
\usepackage{bm}
\usepackage{amsfonts,amssymb,amsmath}
\usepackage[colorlinks,linkcolor={blue},citecolor={blue},urlcolor={blue}]{hyperref}
\usepackage{textcomp}
\usepackage{float}

\usepackage{color}
\usepackage{array}
\usepackage{booktabs}
\usepackage{multirow}
\begin{document}

\title{Inducing chiral superconductivity on honeycomb lattice systems}

\author{Abdulrhman M. Alsharari}
\email{aalsharari@ut.edu.sa}
\affiliation{Department of Physics, University of Tabuk, Tabuk, 71491, SA}
\author{Sergio E. Ulloa}
\affiliation{Department of Physics and Astronomy, and Nanoscale and Quantum
	Phenomena Institute, Ohio University, Athens, Ohio 45701, USA}
\affiliation{Center for Quantum Devices, Niels Bohr Institute, University of Copenhagen, DK-2100 Copenhagen, Denmark}
\affiliation{Center for Nanostructured Graphene, DTU Physics, Technical University of Denmark, DK-2800 Kongens Lyngby, Denmark}
%

\date{\today}

\begin{abstract}	
	Superconductivity in graphene-based systems has recently attracted much attention, as either intrinsic behavior or induced by proximity to a superconductor may lead to interesting topological phases and symmetries of the pairing function.  
A prominent system considers the pairing to have chiral symmetry.  The question arises as to the effect of
possible spin-orbit coupling on the resulting superconducting quasiparticle spectrum.  
Utilizing a Bogolyubov-de Gennes (BdG) Hamiltonian, we explore the interplay of different interaction terms in the system, and their role in generating complex Berry curvatures in the quasiparticle spectrum, as well as non-trivial topological behavior.  
We demonstrate that the topology of the BdG Hamiltonian in these systems may result in the appearance of edge states along the zigzag edges of nanoribbons in the appropriate regime.  
\end{abstract} 

\maketitle

\section{Introduction}

The intriguing properties of topological matter, especially superconductivity, have greatly motivated the ongoing search for such behavior in different material classes \cite{Sato.Topological.2017,Varona}. Among these searches, interest in two-dimensional (2D) materials has been intense \cite{KaneMele}.   Graphene-based 2D systems promise unusual properties and applications in a variety of spintronic designs. Achieving this behavior involves non-trivial properties, such as the quantum anomalous Hall effect, topological phases with complex Berry curvature and associated Chern numbers, helical states that may support Majorana fermion excitations, as well as possible chiral currents along the edges of 2D ribbons \cite{Faye,Volovik.Fra.89,Wang.Topological.2016,Dutreix}. 

The quantum anomalous Hall (QAH) regime has received a great deal of attention at the level of single-particle behavior. An intrinsic magnetization in the system resembles to some extent the effect of a magnetic field on the electronic dynamics with peculiar results \cite{theory,QAHE2011,Fabian2020-2}. In particular, the magnetization breaks time reversal symmetry (TRS) and lifts the spin degeneracy. A quantized charge Hall conductance characterizes this phase, displaying an insulating state in the bulk and edge states at the boundary of a nanoribbon system.  

The addition of superconducting correlations on a QAH system has been predicted to give rise to intriguing new properties and possible Majorana edge states which may be used as robust platforms in quantum computation  \cite{FuKane}.
The induced pairing correlations have naturally a direct impact on the appearance and  specifics of these  properties \cite{Varona}. The symmetries of the resulting pairing channel are closely related to underlying lattice symmetries as well as to the mechanism  
mediating the interaction. 
This is turn produces different quasiparticle spectra with markedly different features. 

In contrast, the proximitized honeycomb lattice of graphene may result in  
broken lattice inversion symmetry and strong spin-orbit coupling that induces an effective Ising (or Zeeman-like) spin field that preserves TRS.\@ 
Such an effect results in the spin projection perpendicular to the plane being a good quantum number but with opposite value in each valley to preserve TRS \cite{SOCTMD,Liu.Three.2013,Fabian2020-1,Alsharari.Mass.2016,Alsharari.Topological.2018}.  Such a situation
could be induced in graphene deposited on a transition metal dichalcogenide substrate, for example \cite{Fabian2020-1,Alsharari.Mass.2016,Alsharari.Topological.2018}. 
Considering the effect of superconducting correlations on such Zeeman-like system is expected to provide non-trivial Chern numbers and complex Berry curvatures that should affect the resulting symmetries of the quasiparticle excitations \cite{Fabian2020-2}.

The classification of superconducting pairing symmetries residing on the honeycomb lattice has been discussed  previously \cite{Varona,Faye,Dutreix,Annica}.  
Although deciding which symmetry dominates remains somewhat controversial, an intriguing possibility of superconductivity induced onto 
proximitized graphene is the appearance of chiral pairing symmetries and their associated broken TRS.\@
For instance, analysis of various symmetry channels in a Hubbard model on a honeycomb lattice shows that a singlet `d+id'-wave is the dominant pairing symmetry  \cite{Ying}. 
Highly doped monolayer graphene has been proposed to exhibit these features, with unusual different properties  \cite{Chern16d+,Nandkishore12Chiral,Gong17Time,Lui13d+,[{For discussion and references on this topic see }]Scherer2020}.

The study of either a QAH or a Zeeman-like sytem in the presence of induced chiral d-wave superconductivity has received little attention, 
and we focus on this issue here. 
Our primary interest is to explore the relationship between the chiral superconducting pairing functions and topological properties of the corresponding Bogoliubov-de Gennes (BdG) Hamiltonian that couples electron and hole subsystems through the superconducting pairing. We explore this effect with a focus on inducing topologically nontrivial character in the proximitized system. We study the quasiparticle (QP) band spectra of the superconducting states, and demonstrate that the topology of the BdG Hamiltonian is different from the constituent subsystems. To this end, we study the topological properties using both the Berry curvature of the QP dispersion and their relation to the appearance of edge states in finite nanoribbon systems.  For suitable chemical potential and superconducting pairing strength, we find the appearance of robust midgap states at zigzag edges, well protected by large excitation gaps and momentum transfer.

In Sec.\ \ref{micmodel} below we describe the low-energy electronic Hamiltonian for the QAH and Zeeman-like regimes of the honeycomb lattice system, and describe the main features and differences of the resulting single-particle band structure.  We then discuss the effect of spatial range on chiral superconducting pairing in Sec.\ \ref{supercmodel}, by considering coupling for nearest and next-nearest-neighbors in the lattice, and what symmetry this entails.  Section \ref{results} explores the main features of the QP spectra for the different regimes and superconducting symmetries, including the appearance of topologically protected edge states on zigzag-terminated nanoribbons. Section \ref{expts} is devoted to a discussion of realistic model parameters and the possibility of realizing the discussed topological phases experimentally. 
Finally, we discuss the importance and the prospects of these findings in the concluding section.

\section{Microscopic Model}
\label{micmodel}

\subsection{Electronic states of proximitized honeycomb lattice}
\label{elecmodel}

The Bogoliubov-de Gennes (BdG) Hamiltonian for superconducting pairing on the graphene honeycomb lattice is described in momentum space as \cite{Alsharari2020} 
\begin{equation}\label{BdG}
\mathcal{H}_{BdG}(k)=
\left( 
\begin{matrix}
\mathcal{H}_{e}(k)-\eta && \Gamma(k)  \\
\Gamma^\dagger(k)   && \eta-s_y\mathcal{H}^{*}_{e}(-k)s_y \\
\end{matrix}
\right) ,
\end{equation}  
where $s_y$ is a Pauli matrix acting on spin space, $\eta$ is the chemical potential of the system, and  $\mathcal{H}_{e}$ is the effective electronic Hamiltonian for proximitized graphene (without superconductivity), which in the basis $\Phi = ({A\uparrow, }{B\uparrow, }{A\downarrow, }{B\downarrow })^T$ can be written as \cite{Alsharari.Mass.2016,Alsharari.Topological.2018}

\begin{widetext}
\begin{equation}\label{matrix-h}
\mathcal{H}_e (k)=
\left( 
\begin{matrix}
A+Z(k) && T(k)             && 0                && R(k)            \\
T^*(k)   &&A+Z(k) && R(-k)            &&  0              \\
0             &&R^*(-k)   &&-A-Z(k) && T(k)            \\
R^*(k)   &&       0          &&   T^*(k) &&-A-Z(k)\\
\end{matrix}
\right) .
\end{equation}
\end{widetext} 
Here, $k$ is the 2D momentum vector, and 
\begin{equation}\label{Matrix}
\begin{split}
T(k) &=  t  \,(1+e^{i k\cdot a_{1}}+e^{i k\cdot a_{2}}), \\
R(k) &=  \frac{R}{3i} (1 + e^{-i\frac{2\pi}{3}}e^{i k\cdot a_{1}} + e^{i\frac{2\pi}{3}}e^{i k\cdot a_{2}}), \\
Z(k) &= \frac{-2 \lambda}{3\sqrt{3}} \left( \sin(k\cdot a_{1})-\sin(k\cdot a_{2})+\sin(k\cdot a_{3}) \right), \\
\end{split}
\end{equation}
where $a_{1}$, $a_{2}$, and $a_{3}$ are the real-space vectors of graphene (defined explicitly in the next section).  The $A$ constant in Eq.\ \eqref{matrix-h} represents the effect of a uniform magnetic or exchange field that induces the system to be in a QAH regime. This term is in clear competition with the effective Zeeman field $Z(k)$ induced by the spin-orbit coupling constant $\lambda$ that embodies the underlying lattice structure and preserves TRS.\@  The $R$ constant represents a possible Rashba spin-orbit coupling, allowed as well by the broken inversion symmetry.  For convenience, we set the nearest neighbor hopping as the unit of energy, $t = 1$, throughout the paper.

We now illustrate the differences between the QAH and Ising spin-orbit honeycomb systems {\em before} the onset of superconductivity.
The electronic Hamiltonian in the  QAH regime, which we will denote as an {\em A-System}, uses $A=0.05$ and $\lambda = 0$.  Similarly, we will look at a {\em Z-System}, where $\lambda = 0.05$ and $A=0$.  In both cases we assume the presence of a Rashba
field with $R=0.05$, which has as main effect to mix different spins, splitting their degeneracy (increasing the Rashba coupling does not produce a transition to a topological phase \cite{KaneMele2005}).
The parameters chosen capture the essential characteristics of each regime; systems with other coupling constant values exhibit the same qualitative behavior presented here. 
We note that other terms in the Hamiltonian may exist in a proximitized graphene system, such as that produced by a difference between A and B sublattices (the `staggered' field) or an intrinsic spin-orbit coupling term. However, we have verified that as long as such parameters remain small, as it is typical in these structures, their inclusion in the model has only a small quantitative effect, with no change in the qualitative behavior. 

\begin{figure}[t]
	\centering
	\includegraphics[width=\linewidth]{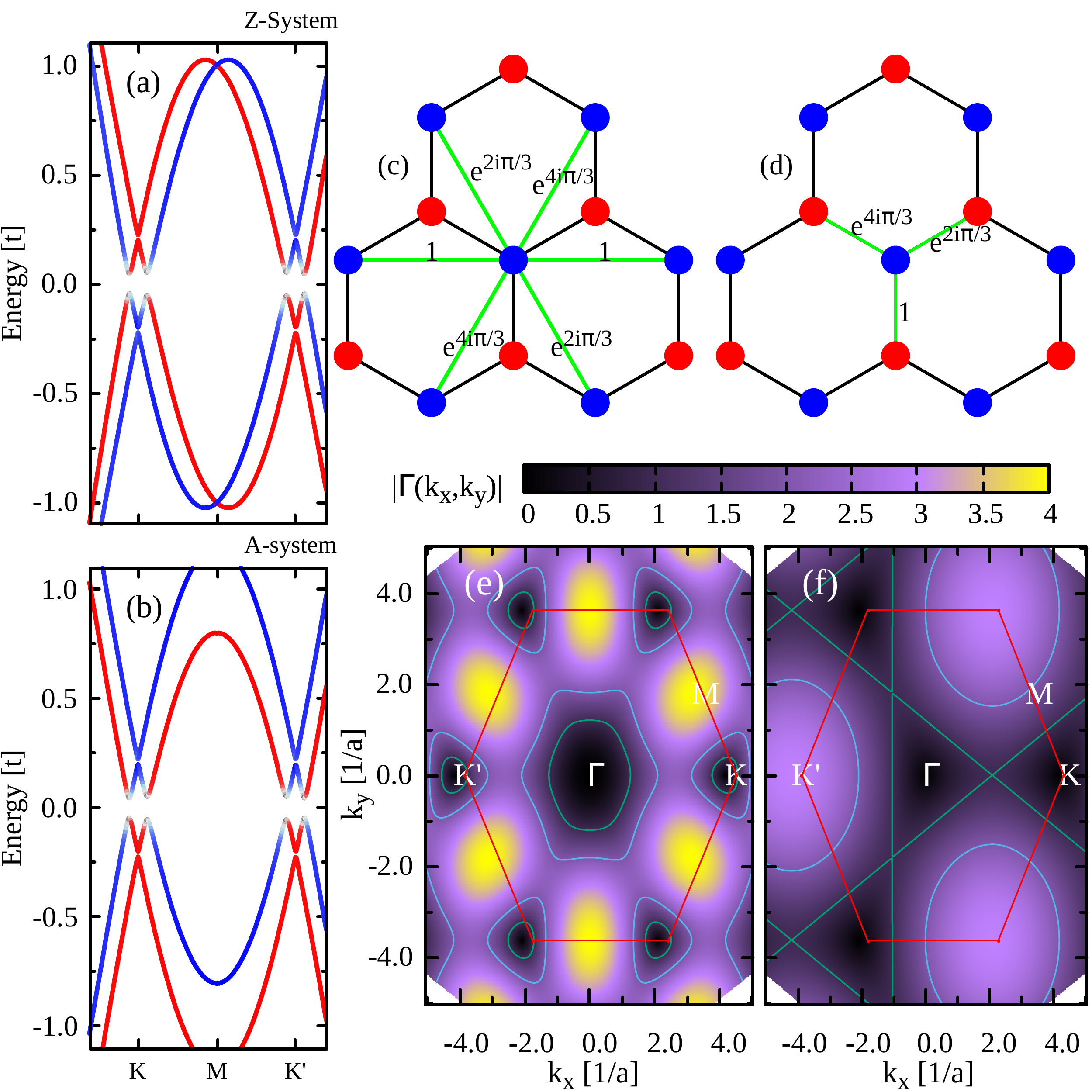}
	\caption{(Color online) Panels (a) and (b) show band structure of the electronic part of the BdG Hamiltonian for two different regimes. 
	(a) A {\em Z-System} is the structure that results in the presence of Ising spin-orbit coupling, here $\lambda=0.05t$.   (b) An {\em A-System} results when a uniform magnetic (or intrinsic exchange) field leads to appearance of QAH effect; here $A=0.05t$. 	
	Panels (c) NNN, and (d) NN, show phase factors in real space introduced by the chiral superconducting pairing function in each case.  Corresponding $|\Gamma(k)|$ function shown in reciprocal space for (e) NNN, and (f) NN.  Note drastically different symmetries in both cases. } %
	\label{Fig1}
\end{figure}

Panels (a) and (b) of Fig.\ \ref{Fig1} show the electronic band structure of both the {\em Z-} and {\em A-Systems}, with their corresponding spin $S_z$-projection, along the K-M-K' path in the Brillouin zone. Notice that K and K' states have the same spin ordering in the {\em A-System} whereas the spin is reversed in the {\em Z-System}. This is understood since QAH breaks TRS, whereas this symmetry is preserved by the Ising spin-orbit coupling in the {\em Z-System}.
As we will see below, the presence of chiral superconducting order parameters makes the BdG Hamiltonian of the {\em Z-System} odd under time-reversal. A chiral gap function breaks TRS and hence allows splitting of bands that were once degenerate.

\subsection{Superconducting pairing function}
\label{supercmodel}

We want to consider a chiral pairing amplitude $\Gamma(k)$ that breaks time-reversal symmetry  in Eq.\ \ref{BdG}. 
Moreover, we want to explore the differences on superconducting QP spectra arising from either nearest neighbor (NN) and next nearest neighbor (NNN) coupling, as the spatial content results in different momentum-space pairing symmetries.
The pairing intensities are accompanied with different variables/factors for different directions, as shown in Fig.\ \ref{Fig1}(c) and (d). As a result, the non-zero elements of $\Gamma(k)$ in Eq.\ \eqref{BdG} for d$_{x^2-y^2}+ i$d$_{xy}$ superconducting order can be expressed as
\begin{equation}\label{OP4}
\begin{split}
\Gamma^{NN}_{A\uparrow,B\downarrow}(k) &=   \gamma \left( e^{-i k\cdot \delta_{1}} + e^{i\frac{2 \pi}{3}}e^{i k\cdot \delta_{2}}  + e^{i\frac{4 \pi}{3}}e^{i k\cdot \delta_{3}}  \right)  ,\\
\Gamma^{NNN}_{A\uparrow,A\downarrow}(k) &= 2\gamma \left(  \cos\left(k_x \right) - \cos\left(\frac{k_x}{2} \right) \cos\left(\frac{\sqrt{3} k_y}{2} \right) \right. \\ & \,\,\,\,\, - i  \left. \sqrt{3} \sin\left(\frac{k_x}{2}\right)\sin\left(\frac{\sqrt{3} k_y}{2} \right)    \right) , 
\end{split}
\end{equation} 
where $k=(k_x,k_y)$, and the graphene lattice constant $a$ is set to 1.  
We have defined the honeycomb vectors connecting nearest neighbors (sublattice A to sublattice B) as
\begin{equation}\label{OP0}
\begin{split}
\delta_1 &= a (0,\frac{1}{\sqrt{3}})\\
\delta_{2,3} &= \frac{a}{2} (\pm 1 ,\frac{1}{\sqrt{3}}).
\end{split}
\end{equation} 
The corresponding lattice vectors connecting sublattice A (or B) to its NNN  are given by $a_1=\delta_2-\delta_3$, $a_2=\delta_1-\delta_3$, and $a_3=a_1-a_2$.

Notice that the symmetries of the singlet pairing function dictate that 
\begin{equation}\label{OP5}
\begin{split}
\Gamma^{NN}_{A\uparrow,B\downarrow}(k) &= \Gamma^{NN}_{B\uparrow,A\downarrow}(-k),\\
\Gamma^{NNN}_{A\uparrow,A\downarrow}(k) &= \Gamma^{NNN}_{A\downarrow,A\uparrow}(-k),\\
\Gamma^{NNN}_{A\uparrow,A\downarrow}(k) &= \Gamma^{NNN}_{B\uparrow,B\downarrow}(k).\\
\end{split}
\end{equation}

Figure \ref{Fig1}(e) and (f) show the $k$-dependent map of the superconducting pairing amplitude for NNN and NN couplings, respectively.  Notice that the d-wave chiral amplitude has nodal points or lines when the pairing amplitude transitions between negative and positive values in momentum space. In both cases, the $\Gamma(k)$ function has at least C$_3$ symmetry with respect to the $\Gamma$ point \cite{Qi}.  However, the symmetry at the K and K' valleys is very different in both pairings.  This would have significant impact on the QP spectra and associated Berry curvature, as will be seen below. 

We study the role of the chemical potential in the QP spectrum as the superconducting pairing strength changes.  We focus on energies close to the Dirac points and keep both $\eta$ (chemical potential) and $\gamma$ (superconducting pairing strength)  relatively small, $\eta$, $\gamma < t/4$.
This focuses on doping in the vicinity of the K valleys.  Notice that heavy doping requires a more complicated structure, with consideration of different regions in momentum space.  For large doping, the topological properties are dependent on the $\Gamma$ and/or M points near  van Hove singularities, as well as possible band renormalization effects \cite{Dutreix,Wang.Topological.2016,Scherer2020}.

\section{Results and discussion}
\label{results}

\subsection{Quasiparticle spectra in different regimes}

As illustrated in Fig.\ \ref{Fig1}(a) and (b), both the {\em A}- and {\em Z-systems} exhibit an insulating gap around the charge neutrality point that is proportional to the relevant structure parameter ($A$, $\lambda$ and $R$).
To understand the effect of the chemical potential and superconducting pairing function on the different phases of the BdG Hamiltonian, Eq.\ \eqref{BdG}, we show a map of the excitation gap (minimal electron-hole excitation, regardless of its momentum position) near the chemical potential (zero excitation energy) level. Notice that the gap can close at one of the symmetry points (K or K') of the Brillouin zone or away from them, and either way appear gapless in the map.

\begin{figure*}[t]
	\centering
	\includegraphics[width=\linewidth]{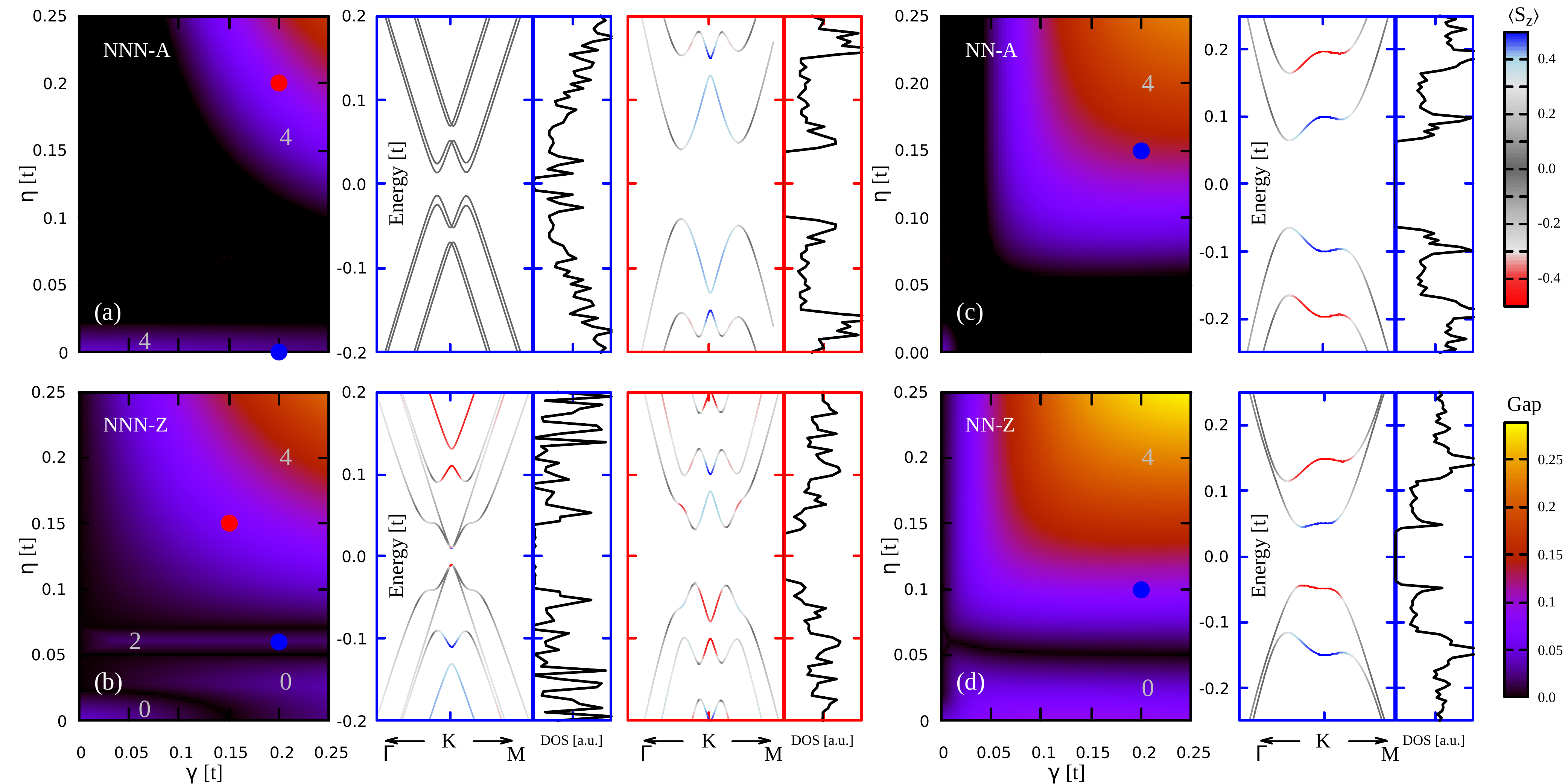}
	\caption{(Color online) Gap phase diagram and QP band structure of systems with NN and NNN chiral coupling. Heat maps show the minimal excitation gap for different systems as function of chemical potential $\eta$ and superconducting coupling $\gamma$; gap scale shown at lower right.  Black regions indicate gapless states that may signal a transition between different topological phases, as discussed in text. 
	The total BdG Chern number is shown in gray for each distinctive phase separated by gapless regions.
	For each system, panels on right display QP spectra and corresponding density of states for selected parameter values (color dots), as indicated by colored frames. Red to blue shading of dispersion curves indicate $S_z$ projection as function of momentum; spin values shown in upper right scale. }
	\label{Fig2}
\end{figure*}

Figure \ref{Fig2} shows the gap behavior in different regimes as function both of chemical potential $\eta$ and superconducting strength 
$\gamma$.  Although the latter is set by the structure and pairing mechanism involved in a given material system, $\eta$ could be modified somewhat by direct doping or even applied gate voltages.
As the gap diagram is symmetric around zero with respect to both $\eta$ and $\gamma$, we plot only the positive quadrant for each case.
For $\gamma=0$, the gapless states reflect the single-particle spectral gap seen in Fig.\ \ref{Fig1}(a) and (b). As either $\eta$ or $\gamma$ increase, however, the gap behavior changes substantially for the different regimes.  
The systems studied show distinct, separate regions with contrasting QP behavior.

In the NNN {\em A-System}, an excitation gap persists even for large $\gamma$, as long as the chemical potential is close to zero, Fig.\ \ref{Fig2}(a).  Increasing $\eta$ leads to gapless phases, as the pairing function nodal structure dominates.  Larger $\eta$ and $\gamma$, however, allows other states to participate in the pairing correlations, which results in extended phases with gap size that is proportional to $\eta$ and $\gamma$, both exceeding the value of $A$ ($=0.05$ here).

To further distinguish/identify the characteristic features of these phases, we plot the QP band structure at selected parameters in these regions in Fig.\ \ref{Fig2}.  We focus on one valley for clarity and show band structure and density of states (DOS) of selected points. 
Each panel to the right of panel Fig.\ \ref{Fig2}(a) corresponds to different $(\eta,\gamma)$ values as indicated by the blue/red dots and associated frames.  The rightmost panel in each case shows the total density of states (DOS) for each spectrum.  The dispersion is shown along the $\Gamma$-K-M path and curves are colored according to their $S_z$ spin projection.
The excitation spectra in all systems here is symmetric around zero, as dictated by particle-hole symmetry. 
As the {\em A-System} is constructed to break TRS, this is carried over after the onset of superconducting correlations, and we find
that the spin of electrons and holes are the same at each K valley, although reversed in opposite valleys about the M point (not shown). 

Figure \ref{Fig2}(b) shows the gap map for the NNN  {\em Z-System}; an excitation gap in the superconducting phase appears immediately as either chemical potential or $\gamma$ increases, with a non-monotonic dependence on $\eta$.  However, beyond $\eta \gtrsim 0.05 = \lambda$, the superconducting phase has a gap that increases with either $\eta$ or $\gamma$.  The non-monotonic gap dependence, especially the appearance of a gap closing at $\eta \simeq \lambda$, suggests that the system goes through a phase transition.  We will analyze this statement further below, as it relates to changes in Berry curvature and Chern numbers.  
As before, the right frames in Fig.\ \ref{Fig2}(b) show typical QP dispersion and DOS for two cases, indicated by the red/blue dots.  Notice how the gap at $(\eta, \gamma)=(0.06, 0.2)$, blue dot, appears right at the K point, while for (0.15, 0.15), red dot, the gap has shifted away,  suggesting band inversion in the QP states.
Notice that since the {\em Z-System} preserves TRS even as the pairing sets in, the spin states of the bands around the zero energy are reversed in the K and K' valleys.

In the case of NN coupling shown in Fig.\ \ref{Fig2}(c) and (d) the scenario is different. 
For NN-coupling in the {\em A-System}, Fig.\ \ref{Fig2}(c) shows that the normal single-particle gap closes for small values of $\eta $ and $\gamma$, while a gapless regime persists for $\eta \lesssim A $ and all values of pairing strength $\gamma$. 
For larger $\eta$, a superconducting gap emerges and grows with both $\eta $ and $\gamma$ values, similar to the {\em Z-System}. This indicates that in these regimes the bands near the Fermi level couple strongly through the pairing potential, opening a sizable  superconducting gap.
The opening large gap as the chemical potential increases can be seen as arising from the overlap of single-particle states with different symmetry. These bands couple through the pairing function and open a broad QP excitation gap.

The {\em Z-System} in \ref{Fig2}(d) exhibits a sizable gap that persists for all non-zero superconducting coupling and small chemical potential, $\eta < \lambda = 0.05$; the gap is associated with the single-particle gap in Fig.\ \ref{Fig1}(b).  In other words, this system exhibits no phase transition for small $\eta$ as the pairing strength increases.  
Notice that even though there are gapless lines, they do not fully enclose/surround a region. One could then continuously change these parameters without having to close the excitation gap. 
As $\eta $ increases beyond $\lambda$, the system closes the single-particle gap and reopens in a superconducting phase with a gap that 
increases monotonically as both $\eta$ and $\gamma$ increase.

The QP band structure behaves differently as well.  The bands around zero excitation energy are nearly fully spin-polarized around the K (or K') valley, whereas bands at higher energies mix. This is a direct consequence of the NN pairing that couples A and B sublattices.
The total density of states acquires sharp peaks around the Fermi level, as well as for energies $\simeq \gamma$, indicating the emergence of low-dispersive bands, as shown on the right panel of \ref{Fig2}(d).

Comparison of the QP dispersions in both regimes further illustrates differences, such as the spin content of the states.  
While the {\em A-System} shows strongly mixed states, the QP bands in the {\em Z-System} are fully polarized, except around zero energy,  where the states mix due to the Ising and Rashba spin-orbit coupling terms.  As we will see below, the superconducting gap that appears on opposite sides of gap closing regimes have different characteristics. 

Since the pairing function $\Gamma(k)$ has nodal points in the first BZ, Fig.\ \ref{Fig1}, shifting $\eta $ to regions in the spectrum with overlapping particle and hole bands leads to gapless phases in Fig.\ \ref{Fig2}. It is also understandable from the symmetry of particle and hole bands that large regions are gapless, even for non-zero coupling parameter $\gamma$.  Lastly, notice that the chiral NNN d-wave coupling appears to have small effect on the bands of the {\em A-System} at small $\eta$ regimes, \textit{i.e.}, the QP band structure does not change much as a function of $\gamma$.

Notice the expected enhancement of the DOS near the superconducting QP gap, generally larger in the {\em Z-System}  than in the {\em A-System}. [The latter is not evident in Fig.\ \ref{Fig2}, as the DOS are enlarged for clarity in the {\em A-Systems} to emphasize qualitative behavior.]  
The larger DOS is accompanied by low band velocity (flatter QP dispersion).

\subsection{Topological properties} 

 \begin{figure}[t]
 	\centering
 	\includegraphics[width=\linewidth]{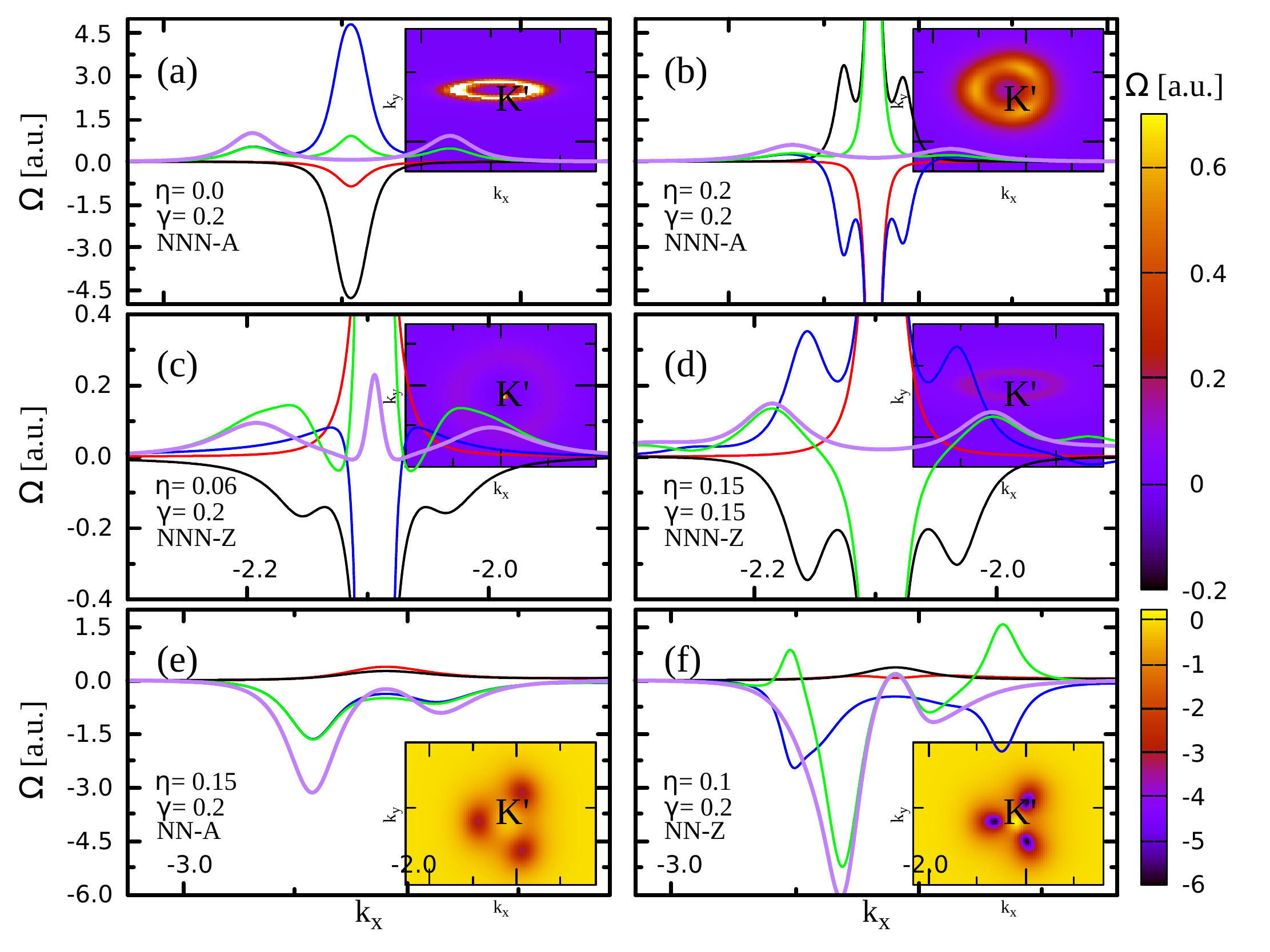}
 	\caption{(Color online) Berry curvature of QP dispersion at K' valley for BdG Hamiltonian at selected parameters corresponding to those in Fig. \ref{Fig2}, as indicated in each legend. The top (middle) panels describe the NNN singlet superconducting {\em A-System} ({\em Z-System}) Berry curvature, whereas the bottom panels describe NN coupling of {\em A-} and {\em Z-Systems}.  Insets show heat maps of the total Berry curvature for holes. Notice circular symmetry of the NNN pairing in both systems, whereas a 3-fold symmetry is evident for NN coupling. }
 	\label{Fig3}
 \end{figure}

We analyze further the topological properties of these systems using the Berry curvature of the QP dispersions.
Differences in Berry curvature  would help us distinguish their physical features, as it has implications on the linear response and overall 
spintronic and valleytronic behavior \cite{berry}. 
One can analyze the Berry curvature of the hole bands per valley, $\Omega_{n}(\boldsymbol{k})$, using \cite{berry}
\begin{eqnarray}\label{berry}
\Omega_{n}(\boldsymbol{k})&=&-\sum_{n'\neq n}\frac{2 \, {\rm Im}\langle\Psi_{n'\boldsymbol{k}}| v_x |\Psi_{n\boldsymbol{k}}\rangle\langle\Psi_{n\boldsymbol{k}}| v_y|\Psi_{n'\boldsymbol{k}}\rangle}{(\epsilon_n-\epsilon_{n'})^2}, \nonumber \\
\end{eqnarray}
where ${v}_x ({v}_y)$ is the velocity operator along the $x(y)$ direction, and $n, n'$ are band numbers \cite{Alsharari.Mass.2016,berry,TI-SC-review}.
The resulting Berry curvature for each of the hole QP bands for the systems 
in Fig.\ \ref{Fig2}, as well as the total hole curvature near the valley at $K'=(-4\pi/3a,0)$ are displayed in Fig.\ \ref{Fig3}. 
The total Berry curvature $\Omega_{tot}$ shows clear changes in structure, associated with each gap closing event and change in the topological character of the bands seen in Fig.\ \ref{Fig2}.
As the BdG Hamiltonian of these systems has particle-hole symmetry by construction, the curvature plots for the other valley are simply related as $\Omega_{tot}$(K-valley$) = \Omega_{tot}$(K'-valley).  

The total hole Berry curvature in the NNN phases is circularly symmetric, reflecting the symmetry of the 
corresponding superconducting pairing functions near the K valleys. In contrast, the three-fold symmetry of the NN pairing function is reflected in the total Berry curvature for those structures, both in the {\em A-} or {\em Z-System}, as seen in Fig.\ \ref{Fig3}(e) and (f).

As stated earlier, at this level of doping the physics of the honeycomb lattice system is controlled by the two nonequivalent valleys, with identical Berry curvature content.   
One can calculate the Chern number per valley and double it, or equivalently define the total Chern number by integrating over the entire Brillouin zone, 
\begin{eqnarray}\label{chern}
C_{BdG}&=&\dfrac{\pi}{2} \int_{BZ} d^2\boldsymbol{k} \, \Omega(\boldsymbol{k}) \, . 
\end{eqnarray}
We notice that in systems of low-doping graphene, the BdG Chern number changes by even numbers across topological phase transitions. This typically indicates the appearance (or suppression) of pairs of propagating edge states whenever a gap closing transition occurs. 
\footnote{The topological characteristics of models that break TRS while preserving particle-hole symmetry, such as the ones we have here, belong to class D \cite{TI-SC-review}.} 
In the systems we study, despite the gap-closing transitions in the NNN {\em A-System}, we find that $C_{BdG}=4$ remains unchanged for all gapped phases. 
This is the result of the single-particle QAH regime having a Chern number of 2, which doubles with the inclusion of the hole sector.
After the gap closing topological transition in Fig.\ \ref{Fig2} one still gets $C_{BdG}=4$ but with a different nature, as we explain below in the corresponding edge state analysis. 

The NNN {\em Z-System} shows $C_{BdG}=0$ for small $\gamma$ values, which then changes to $C_{BdG}=2$ at higher $\eta$ over a small window, and finally goes to $C_{BdG}=4$ after gap opening at higher $\eta$.
Similarly, both the NN {\em A-} and {\em Z-Systems} show a BdG Chern number of 4 after the gap closing transition.

One finds in the {\em Z-System} that the gapped region at low $\eta$ yields zero Chern number--in contrast to the {\em A-System}--as expected from the single-particle picture that respects TRS; the chiral superconducting parameter seems to produce an effective weak electron-hole coupling.

\subsection{Edge states}

The Chern number in single-particle systems represents the number of chiral fermionic edge states \cite{Xiao2007}. Similarly, the 
BdG Chern number indicates the number of chiral edge states in a finite size superconducting system \cite{Fabian2020-2}.   To better identify the topology of these systems, we have studied the corresponding edge states in a finite ribbon with edges along the zigzag direction.
    
It is known that a QAH system with onsite superconducting singlet pairing yields a topological phase with edge states in a finite size system that cross the excitation gap \cite{Wang.Topological.2016}. 
An interesting open question is whether the topological phase can be destroyed by the introduction of chiral pairing interactions.  
As we will see, these systems exhibit edge states in different regimes.

\begin{figure}[t]
	\centering
	\includegraphics[width=\linewidth]{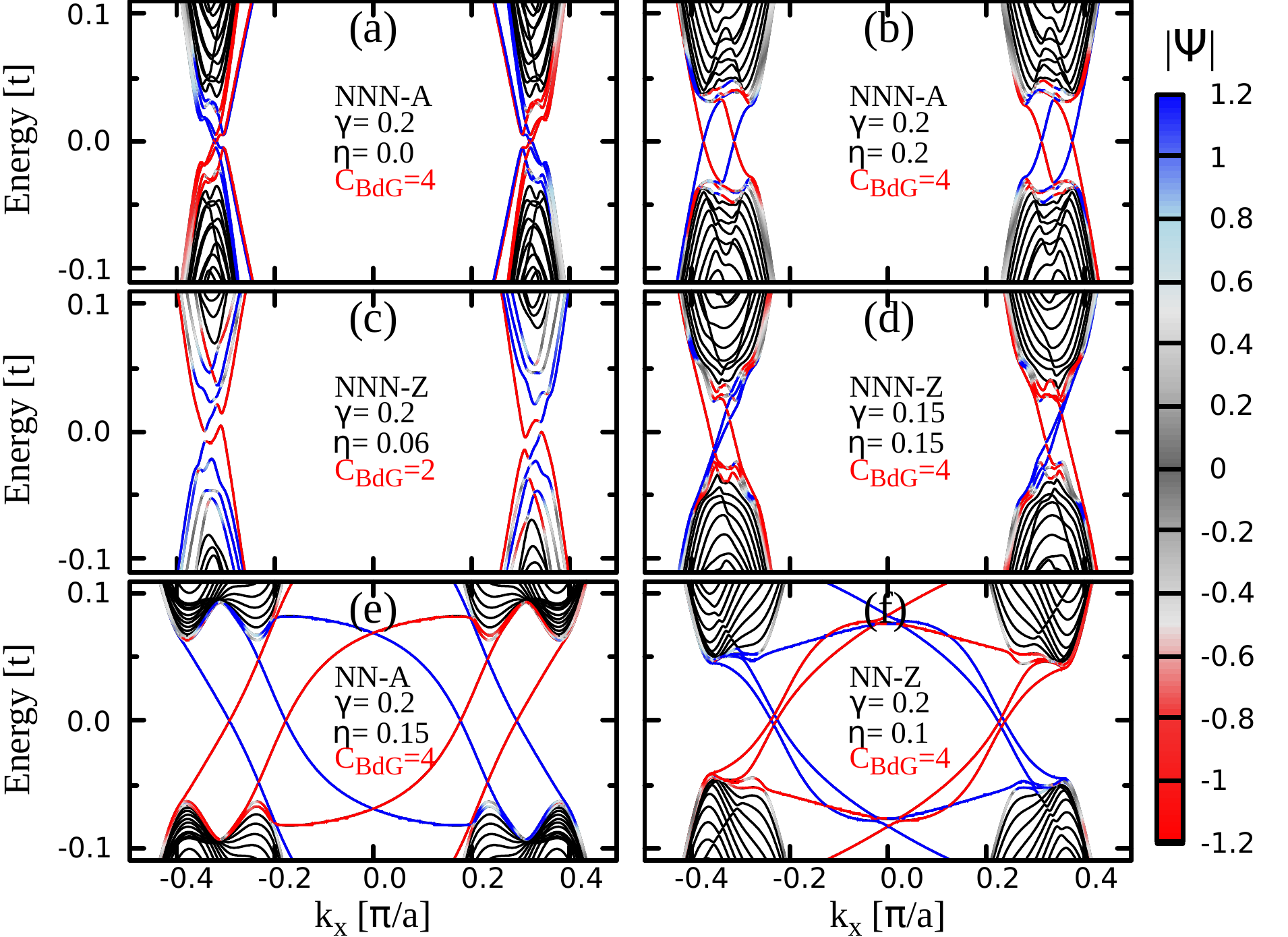}
	\caption{(Color online) Edge states of BdG Hamiltonian at selected superconducting coupling and chemical potential values 
	for the {\em A-} and {\em Z-Systems}. Top panels (a) and (b) are for NNN {\em A-Systems}; (c) and (d) NNN {\em Z-Systems}. 
	Bottom panels show edge states of (e) {\em A-System} and (f) {\em Z-System} for NN coupling. The corresponding $\eta$ and $\gamma$ values are shown in each panel, with corresponding band structures for phases in Fig.\ \ref{Fig2}. 
	Color of each curve indicates spatial distribution of states across the nanoribbon;  red and blue indicate states located at opposite edges, whereas gray indicates the `bulk' middle of the nanoribbon. The BdG Chern number for holes of each phase is also shown, indicating the number of propagating states on each side of the ribbon.}
	\label{Fig4}
\end{figure}

In the NNN pairing for the {\em A-System}, the region of the phase diagram in Fig.\ \ref{Fig2}(a) near $\eta \simeq 0$ has small gaps and show no well-developed midgap edge states.
This is shown in Fig.\ \ref{Fig4}(a), where the QP dispersions for a zigzag nanoribbon with $(\eta, \gamma) = (0,0.2)$ shows a small gap and edge states (as indicated by color, blue = rightmost, red = leftmost) that promptly hybridize with the bulk states.
Well-defined edge states appear after the first gap-closing transition in the system, illustrated in Fig.\ \ref{Fig4}(b) for $(\eta, \gamma) = (0.2, 0.2)$. The edge states connect particle and hole bands as two right- and left-mover modes localized around each of the K and K' points. 
The existence of edge modes in the NNN {\em A-System} does not depend on the specific coupling parameter set (i.e., $\gamma$, $\eta$, $R$, and $A$), once the system is in the upper right sector of Fig.\ \ref{Fig2}(a), which highlights their topological origin. 
The modes are robust and propagate in opposite direction on alternate edges of the nanoribbon, overlapping spatially without coupling.  We emphasize that this results in four (two per valley) co-propagating modes on each side of the ribbon.  

The states change drastically for the NNN {\em Z-System}. Although edge-less for most of the phases shown here, this regime shows a complex structure of edge states for large $\gamma$ values, as seen in Fig.\ \ref{Fig4}(c) and (d). The small gap in (c) is due mostly to the  presence of Rashba coupling.  It is then expected that this structure would not be as robust against perturbation, and that its edge states (either 2 or 4) would be easily perturbed/gapped in the presence of impurities or other perturbations. 

A more complicated edge state configuration is found in cases of NN pairing. In both {\em A-} and {\em Z-Systems}, edge states are present, dispersing across the gap and connecting the two non-equivalent valleys, before eventually merging with the bulk bands. 
The edge states are well protected by the bulk excitation gap, which makes them robust against perturbations.  The phase appears to be topological for large enough values of $\eta$ and $\gamma$ that drive the system through the gap-closing phase transition. 

There are four edge states crossing the gap, with wave-functions localized at the left (right) boundaries of the nanoribbon, as shown by the red (blue) colors in Fig.\ \ref{Fig4}(e) and (f), i.e., there are four modes propagating on the same edge and the same direction.  These states are protected by the large momentum and space separation, as well as sizable gap.

\subsection{Experimentally feasible systems}
\label{expts}

Experimental realization of chiral topological superconductors is actively pursued. New opportunities to study and observe topological superconductivity have been identified in different material systems, especially in 2D crystals.
Localized Majorana states exist at the boundaries of topological superconductors, and can be created by placing chains or islands of magnetic atoms on an s-wave superconductor \cite{SCAdatom,Pascal2020}.  Realization of superconductivity by proximity using hybrid systems, combining magnetism and strong spin orbit coupling, creates novel quasi-particle excitations that open new possibilities for the realization of topological superconductor phases in experiments \cite{Exp1,Exp2,Exp3,Exp5}.
The ability to control and tune different phases can be experimentally achieved, allowing their characterization through measurements such as magneto-resistance and related transport probes.

Superconducting states at low electron doping are observed below 1 K in WTe$_2$ monolayers \cite{SCinTI}. Similarly, superconductivity in high-quality 2M-WS$_2$ single crystals has been reported by several groups experimentally \cite{Other2019,2MTMDSc}. This unsual material is predicted to exhibit topological nodeless superconductivity, with probable Majorana states bound to vortex cores. 
Similarly promising chiral-triplet superconductivity is predicted in heavy-fermion dichalcogenide UTe$_2$ \cite{UTe}. Possible topological superconductivity is predicted theoretically in other 2D transition metal dichalcogenides, as well as other correlated 2D materials \cite{Christopher}.

Hybrid systems such as WSe$_2$/bilayer-graphene/ WSe$_2$ \cite{Island2019} have been shown to exhibit helical edge states arising from the strong spin-orbit orbit under proper gate and displacement field conditions.  As bilayer graphene can be made superconducting, this arrangement may result in unusual phases with very interesting properties, similar to those we discuss here.  
Our estimates suggest reaching doping levels $\eta \gtrsim 0.1t \simeq 0.3$ eV, which might be possible with electrolyte-gating techniques.

These works and others continue to open novel possibilities for topological superconductivity in van der Waals materials with spin-orbit coupling, representing promising platforms to realize and probe non-trivial phases as the ones we have discussed.

\section{Conclusions}
Starting from graphene-like honeycomb lattice models, we study systems with singlet chiral superconductivity to explore the interplay between different topological phases and superconducting range. The NNN correlations result in pairing channels that are rotationally symmetric in momentum space near each valley, whereas 
for NN correlations the symmetry is three-fold.  
The chiral superconducting pairing function studied here is the result of  the complex orbital combination of d$_{xy}$ and d$_{x^2-y^2}$. This combination breaks time reversal symmetry and is shown to be the preferred channel in some graphene models, including those with strong doping (or intercalation).

Tuning the chemical potential produces possible phase transitions that are accompanied by gapless regimes.  
The topological aspects of these phases are identified and studied by looking into the formation of edge states in finite size systems, 
as well as their connection to the Berry curvature of the quasi-particle spectrum and associated Chern numbers in the 2D bulk.
Across topological transition points, the system changes between trivial and non-trivial phases, with different features of edge states in different parameter ranges.  
We trust the present work  would lead to further interest and work in these systems. The general nature of the model suggests that it can be applicable, with suitable variations, to a wide class of honeycomb lattice systems that involve different proximity effects.
 
\section*{Acknowledgments}
SEU appreciates the hospitality at CNG/DTU and NBI/KU, as well as the support of Nordea Fonden and the Otto Moensted Visiting Professorship Program.
 
 \bibliography{cites}
 \bibliographystyle{apsrev4-2}

\end{document}